# A Reusable Four-Port S-Parameter to Link-Level Signal-Integrity Analysis Framework for High-Speed Detector Interconnects

**D M S Sultan[a,1], D. Bortoletto[a], R. Plackett[a], A.E. McDougall[a], A.S. Rotelli[a], A.J.A. Knight[a], J. Vossebeld[b]**

[a] *Oxford Physics Microstructure Detector Laboratory, University of Oxford, OX1 3PU, United Kingdom*
[b] *Department of Physics, University of Liverpool, L69 7ZE, United Kingdom*

**Abstract:** A reusable MATLAB signal-integrity (SI) framework is presented that converts compatible four-port S-parameter data, measured by VNA or obtained from electromagnetic simulation, into traceable link-level evidence rather than a single loss metric. Its contribution is the checked, traceable automation of established SI operations while keeping source data, model assumptions, analytical projections, and finite-record observations distinct. The framework is demonstrated on the four 1.25 Gbps differential routes (DP1-DP4) of the PPCB-1347-MuPix11 probe card using PTSL CST Microwave 3D-Solver-derived four-port S-parameters and a virtual time-domain solver. The automated pipeline preflights file structures, performs a power-normalized mixed-mode transformation, applies route-length-aware loss decomposition, constructs a causally loaded channel model, and propagates full PRBS-31 sequences into eye-diagram, conditional-BER, and 8b10b-coded-link analyses. At the 1.25 Gbps data rate (Nyquist 0.625 GHz), DP1–DP4 exhibit differential insertion loss (SDD21) from −0.350 to −0.300 dB, differential-to-common conversion from −31.038 to −28.236 dBc, and modeled FEB-input eye openings from 0.586 to 0.588 V. The comparison shows that path length alone is not an adequate SI ranking variable: DP3 has the lowest Nyquist insertion loss, DP1 the strongest differential-to-common isolation, and DP4 the largest modeled eye. All analytical BER values remain below the reporting floor and therefore do not support a BER ranking. By preserving the distinction between route-dependent waveform behavior, model projections, and finite-record observations, the framework provides an extensible basis for comparative high-speed-interconnect SI analysis from design review through calibrated VNA measurement interpretation.

**Keywords:** Signal integrity, S-parameters, mixed-mode conversion, high-speed link modeling, probe card, MuPix11, Mu3e experiment, eye diagram, Bit Error Rate (BER), 8b10b encoding, loss decomposition.

[1] Corresponding Author
Email: dms.sultan@physics.ox.ac.uk

## Contents

## 1. Introduction

At the Paul Scherrer Institute, Mu3e investigates the charged-lepton-flavor-violating decay $\mu^+ \rightarrow e^+ e^- e^+$. Reconstructing its low-momentum electron and positron tracks demands a low-mass detector with fine spatial resolution and sustained operation at high particle rates. Mu3e meets these requirements with thin MuPix high-voltage monolithic active pixel sensors (HV-MAP), which integrate the sensing region, analog front end, digital hit processing, and data transmission on a single chip [1].

On the transmitting side of the link considered here, the MuPix digital periphery collects hit addresses and timestamps, forms eight-bit data and control words, applies 8b10b coding, and serializes the resulting ten-bit symbols onto 1.25 Gbps differential outputs. MuPix11 chips established this streaming architecture and the 1.25 Gbps line rate [1]. Its reference-clock/PLL and synchronous reset (SYNCRES) operate at 125 MHz; the serializer runs at 800 MHz; and the slow control bus (SIN) runs at 10-40 MHz [2]. The PPCB-1347 probe card includes four MuPix11 high-speed transmitter pairs, denoted DP1-DP4, together with the lower-rate clock, slow-control, and reset paths.

At the receiving end of the detector link, the Mu3e Frontend Board (FEB) is centered on an Intel/Altera Arria V FPGA. Its clock network distributes 125 MHz detector and FPGA clocks and provides dedicated transceiver reference clocks; the FPGA receives and processes MuPix data before the FEB packetizes the event stream for onward optical transfer to the switching board [3,4]. The present study stops at the electrical FEB input. Accordingly, the 100 Ω differential termination, 0.35 pF input capacitance, 2 mV RMS voltage noise, and 2 ps RMS aperture jitter used below are explicit receiver-model assumptions, not measured Arria V input specifications.

The motivation is to determine whether the shared PTSL probe-card layout preserves the MuPix-to-FEB signal margin before hardware is available for a complete end-to-end qualification. PTSL supplied route geometry, four-port Computer Simulation Technology (CST) Touchstone data, and simulated time-domain reflectometry (TDR); the TDR indicates a nominal 100 Ω differential trace with localized departures at the probe head, vias, and connector transitions [5]. Raw S-parameters alone do not quantify polarity, mixed-mode conversion, data-dependent inter-symbol interference, sampling-phase margin, or coded-link error consequences. The Oxford MATLAB framework therefore keeps DP1-DP4 traceable to their individual files and physical lengths, verifies port pairing, converts each four-port response to mixed mode, constructs a causal loaded channel, propagates a complete PRBS waveform, and separates eye, conditional-BER, and 8b10b checker projections from finite-record observations. Cross-route statistics are formed only after every route has completed the same independent analysis; no S-parameter matrices are averaged, and no comparison with another ladder or measurement project is made. The framework contribution is this common, traceable chain of checked operations and declared assumptions, rather than a new electromagnetic solver or a claim of hardware validation.

## 2. MATLAB-based analysis framework

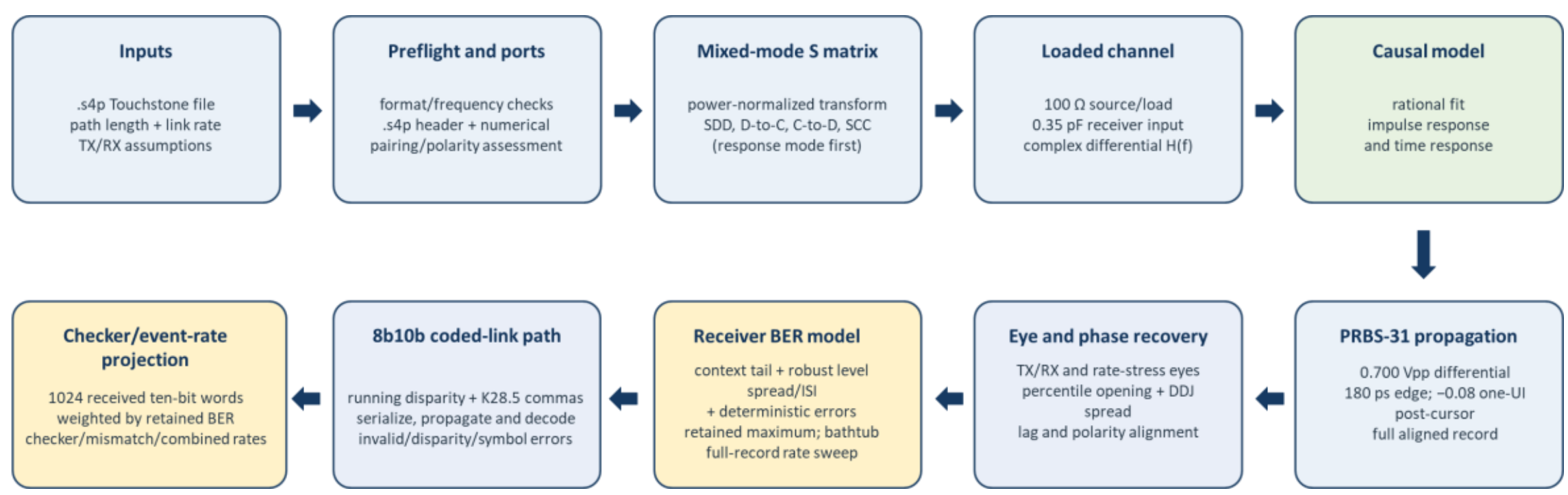

Figure 1. MATLAB analysis sequence from four-port Touchstone input to receiver BER and 8b10b checker/event-rate outputs. Error-rate projections are separated from observed finite-sample counts and hardware measurements.

Figure 1 summarizes the per-route MATLAB implementation. Each CST Touchstone file is first checked for frequency coverage, storage convention, and physical port pairing; the CST header assignment is cross-checked numerically before a power-normalized single-ended-to-mixed-mode transformation is applied. The response-mode-first matrices provide differential transfer, differential-to-common conversion, common-to-differential conversion, and common-mode transfer. A loaded differential channel is then formed with the declared 100 Ω source and receiver impedances, and 0.35 pF receiver input, converted to a causal rational model, and excited by the full aligned PRBS-31 waveform using the stated 0.700 $V_{pp}$ differential launch, 180 ps edge, and one-UI post-cursor. Eye opening, crossing spread, polarity, lag, and sampling phase are recovered from the propagated waveform. The retained receiver BER is the maximum of the context-resolved Gaussian tail, the robust level-spread/ISI term, and any deterministic decision errors; the same complete decision record is used for the bathtub and data-rate sweep. Finally, the running-disparity 8b10b encoder/decoder path and exhaustive immediate ten-bit-word enumeration convert the raw BER model into separated checker, valid-symbol-mismatch, undetected-mismatch, and combined event-rate projections. Observed finite-record counts, model projections, and the $10^{-15}$ reporting floor remain distinct in the Oxford SI framework outputs.

### 2.1. Mixed-mode representation

Power-normalized sum and difference waves separate desired differential transfer from deterministic modal conversion. The transformation is algebraic and contains no stochastic source; common-mode noise can be calculated only when both a C-to-D transfer function and a source spectrum are supplied [6,7].

$$a_d = \frac{a_1 - a_2}{\sqrt{2}}, a_c = \frac{a_1 + a_2}{\sqrt{2}} \tag{2.1}$$

$$\begin{bmatrix} b_d \\ b_c \end{bmatrix} = \begin{bmatrix} S_{DD} & S_{CD} \\ S_{DC} & S_{CC} \end{bmatrix} \begin{bmatrix} a_d \\ a_c \end{bmatrix} \tag{2.2}$$

Here, excitation-to-response denotes: SDD is differential-to-differential transmission, SDC is differential excitation observed as common mode, SCD is common excitation observed as differential mode, and SCC is common-to-common transmission. MATLAB's internal array indices place response mode before excitation mode; the manuscript follows the displayed physical direction and spells it out whenever ambiguity is possible.

### 2.2. Conductor, dielectric, and higher-order loss

The smooth attenuation of a high-speed data interconnect is represented by physically motivated frequency bases over the declared fit band. For a good conductor, current crowds into a skin depth that decreases as the inverse square root of frequency, while surface resistance rises as the square root of frequency [8].

$$\delta(f) = \frac{1}{\sqrt{\pi f \mu \sigma}},\ R_s(f) = \sqrt{\frac{\pi f \mu}{\sigma}} \tag{2.3}$$

A causal dielectric term is approximately linear in frequency when effective permittivity and loss tangent vary slowly. Residual curvature is represented by non-negative quadratic and quartic bases. The dielectric-loss and higher-order terms are quantitative contributions of the declared physics-motivated MATLAB loss model: together with the conductor term and whole-path intercept, they reconstruct the selected SDD21 response under the stated path-length, frequency-band, and dispersion assumptions. They are not separate direct measurements of material loss tangent or radiated power. The Svensson-Djordjevic causal dielectric-dispersion shape is used over the selected fit range [9].

$$IL_{fit}(f) = A_{fixture} + L[a_c\sqrt{f} + a_d f + a_2 f^2 + a_4 f^4] \tag{2.4}$$

Here, $a_c\sqrt{f}$ denotes conductor loss, $a_d f$ points dielectric loss, and $a_2 f^2 + a_4 f^4$ quantifies the higher-order residual-curvature proxy. The non-negative fit is evaluated over 0.01–0.80 GHz. The 0.01 GHz lower limit retains the measured low-frequency response needed to anchor the whole-path offset; the 0.80 GHz upper limit includes the 0.625 GHz Nyquist operating point with a guard interval. Within this finite band, the sqrt(f), f, $f^2$, and $f^4$ bases can describe similar smooth attenuation trends, so their fitted coefficients are correlated, and the decomposition is not unique. Nevertheless, each reported term is a quantitative contribution of this declared model; together with the whole-path intercept, the terms reconstruct the smooth SDD21 response. A coefficient of zero means only that this constrained fit did not require that basis over 0.01–0.80 GHz; it does not imply that the corresponding physical mechanism is absent. A non-zero higher-order contribution summarizes residual curvature and cannot by itself identify radiated power, a via transition, or a return-path discontinuity. Additional coupon, material, or de-embedded fixture information would be needed only to identify those physical mechanisms separately.

## 2.3. Mixed-mode conversion and disturbance transfer

In a finished PCB, layouts are never perfectly symmetric (due to glass weave skew, trace length mismatches, asymmetrical via patterns, or bent traces); the channel converts energy between modes. The differential noise variance $\sigma_{d,\mathrm{CM}}^2$ can be evaluated using the following relation:

$$\sigma_{d,\mathrm{CM}}^2 = \int_0^B |H_{C\to D}(f)|^2 S_{v,\mathrm{CM}}(f)df \tag{2.5}$$

Here, B is the receiver bandwidth, $H_{C\to D}(f)$ refers to the channel's conversion efficiency at frequency, and $S_{v,\mathrm{CM}}(f)$ denotes the voltage noise spectral density of the common-mode disturbance (aggressors, power-supply ripple, crosstalk). However, the results in Figure 4(b) for the channel were analyzed using S-parameters from the pre-production simulation and are therefore completely bounded by the statistical error limit assigned to the standard model used, rather than by real termination-imbalance noise or receiver common-mode rejection versus frequency.

To avoid double counting, the deterministic D→C/C→D conversion stays in Figure 4(b). Deterministic ISI stays in the context-specific sampled voltage. Only the declared 2 mV RMS receiver noise and the equivalent voltage uncertainty from the 2 ps aperture jitter, $\sigma_{v,jitter} = \left|\frac{dv}{dt}\right|\sigma_t$, enter the conditional Gaussian variance. The diagnostic level-spread proxy is excluded from production BER because treating deterministic multimodal ISI as independent Gaussian noise creates false BER structure.

## 2.4. Causal channel and waveform construction

The selected differential transfer function is interpolated onto the simulation grid with magnitude and unwrapped phase treated coherently. A causal impulse response is formed, and the source waveform is convolved with that impulse response. Any time shift used for display is common across comparisons and does not create eye-opening.

In MATLAB, the selected loaded differential transfer function H(f) is sampled on a strictly increasing frequency grid and checked for finite complex values. The declared 100 Ω source and 100 Ω receiver impedances and the 0.35 pF receiver input are included before fitting, so mismatch and receiver-capacitance effects remain in both gain and phase. MATLAB rational approximates these complex frequency samples with a finite stable pole-residue linear-time-invariant (LTI) transfer function at the configured fitting tolerance. This is a compact causal interpolation of the S-parameter response, not an additional physical loss model. Its impulse response h(t), together with the physical bulk delay, becomes the route-specific channel object used in the subsequent waveform calculations; non-causal pre-cursor energy from finite-band interpolation is discarded rather than interpreted as channel response.

The transmitter record is constructed from the complete aligned PRBS-31 sequence. Each bit is mapped to a bipolar differential level; the configured one-UI post-cursor is added to the preceding-symbol contribution; and a causal first-order edge model produces the 180 ps transition. The resulting launch waveform x(t) is propagated through the causal loaded-channel impulse response h(t). No magnitude-only filter or eye-specific channel is used, so the received waveform y(t) retains the same frequency-dependent attenuation, group delay, reflection structure, and deterministic inter-symbol interference as the S-parameter-derived channel:

$$y(t) = \int x(\tau)h(t-\tau)d\tau \tag{2.6}$$

In equation (2.6), x(t) is the differential launch waveform, h(t) is the causal impulse response of the loaded S-parameter channel, $\tau$ is the integration-time variable, and y(t) is the resulting differential waveform at the receiver input. The convolution therefore applies the route-specific channel memory to the complete transmitted record.

Receiver polarity is corrected before the decision statistics are formed. MATLAB first evaluates the signed dot product of an initial transmitted and received waveform segment; a negative value causes one inversion of the received differential waveform, corresponding to a swapped pair or 180° receiver-polarity correction. At the decision stage, the known PRBS sequence is then aligned over the permitted whole-symbol lags, and the matching routine returns the decision polarity used for the selected samples and slopes. That recovered polarity and lag are consistently retained across phase recovery, conditional BER, bathtub scans, rate sweeps, and the 8b10b decoder. A later circular shift may center the plotted eye, but it is display-only and does not change the decision samples or the reported BER.

### 2.5. Eye metrics

Samples or time durations of a single bit are folded into the unit interval (UI). Rather than measuring simple peak-to-peak voltage, the eye height ($H_{\text{eye}}$) is evaluated at a specific sampling phase $\varphi$ (usually the center of the eye width). The absolute minimum '1' and absolute maximum '0', a single isolated noise glitch or transient outlier would make the entire eye look completely closed. Using statistical percentiles filters out rare outliers, giving a stable, statistically robust measurement of signal opening; thus, the framework applied the lower logical-one percentile $P_1(\varphi)$ minus the upper logical-zero percentile $P_0(\varphi)$ at the selected phase. Inter-symbol interference, or deterministic jitter (DDJ), is extracted from the context-dependent crossing spread to quantify how previous bit patterns affect the timing of current bit transitions. Total eye distortion is computed from the TX-to-RX loss of vertical opening [10].

$$H_{\text{eye}}(\varphi) = P_1(\varphi) - P_0(\varphi) \tag{2.7}$$

### 2.6. Receiver and conditional BER model

The statistical model used at the RX calculates the total Gaussian noise standard deviation at the receiver's sampling slicer for a specific bit context $k$. Because of Inter-Symbol Interference (ISI), the voltage at the receiver depends heavily on the preceding sequence of bits (e.g., the intended bits ($b_k$) 0-0-0-1 results in a different sampling voltage than 1-0-1-0). Deterministic ISI is kept directly inside the noiseless, deterministic voltage level predicted by the channel model, $v_k$. The model does not approximate channel attenuation or ISI as "random Gaussian noise." It isolates true random noise separately. The equation calculates total effective voltage noise by combining two independent Gaussian noise sources in RSS (root-sum-squares) fashion.

$$\sigma_k = \sqrt{\sigma_v^2 + (s_k \sigma_t)^2} \tag{2.8}$$

Here, $\sigma_v$ is the pure amplitude noise generated by the Front-End Board (FEB) or receiver electronics (e.g., thermal and active circuit noise). A flat 2 mV RMS value is used in this context. $s_k \sigma_t$ refers to the aperture jitter—the timing uncertainty in the receiver's sampling clock, 2 ps RMS. Since the slicer evaluates voltage rather than time, timing uncertainty $\sigma_t$ is converted into an equivalent voltage noise by multiplying it by the waveform's local slope, $s_k$.

$$\Delta v \approx \frac{dv}{dt} \cdot \Delta t = s_k \sigma_t \tag{2.9}$$

At the eye center ($s_k \approx 0$), the signal curve is flat at the top or bottom rail. A 2 ps timing shift to the left or right results in almost no change in voltage. Total noise is thus dominated purely by 2 mV receiver noise. At near signal transitions ($s_k \gg 0$), the waveform is steep (high $\frac{dv}{dt}$). A 2 ps jitter shift translates into a large voltage fluctuation, causing the jitter-induced term, $s_k\sigma_t$, to dominate total noise.

Instead of using a simple average that may blur everything together, the following equation is used to evaluate the bit error rate (BER) context, assessing the dependence of error probability on the bit pattern. For example, a bit pattern of '0-0-0-1' gets plenty of time to discharge and contributes a large voltage level for the ending '1' transition. In contrast, a bit pattern of '1-0-1-0' toggles so quickly that the trace capacitance never fully charges, resulting in the 1 being much lower in voltage.

$$\mathrm{BER_{context}} = \frac{1}{N}\sum_{k=1}^{N} Q(\frac{M_k}{\sigma_k}) \qquad (2.10)$$

Here, $M_k$ is the distance between the deterministic signal voltage level, $v_k$, and the decision threshold $v_{th}$ for context $k$. Q(x) is the tail probability of a standard normal distribution. It measures the probability that random Gaussian noise will exceed the margin $M_k$, and flip a 1 to a 0 (or vice versa). Using the $\mathrm{BER_{context}}$ equation engine, the BER bath plot is also evaluated for each channel to estimate Clock and Data Recovery (CDR) phase-drift tolerance.

The $10^{-15}$ value is the declared credibility boundary for the BER model used in the analysis. A one-sided Gaussian probability of $10^{-15}$ corresponds to roughly an eight-sigma tail. At that distance, even a few-percent error in the assumed standard deviation or a small unmodeled offset changes the margin, noise RMS, jitter distribution, or threshold, producing orders-of-magnitude changes in Q-function tails. The PTSL-provided S-parameters do not contain measured receiver-noise distributions, supply noise, crosstalk, package nonlinearity, CDR behavior, burst errors, temperature/process variation, or calibrated common-mode interference. Reporting more digits would create false precision. Zero deterministic errors over N decisions instead support the much weaker finite-sample confidence bound shown below.

$$BER_{upper} = \frac{-ln(1-C)}{N} \qquad (2.11)$$

### 2.7. 8b10b encoder, serializer, decoder, and confidence model

The MATLAB coded-link model represents the MuPix transmitter and FEB receiver at the ten-bit-symbol level. It accepts eight-bit data or control inputs, applies the running-disparity 8b10b mapping, serializes the resulting symbols, and decodes the received sequence after channel propagation. The model checks legal-symbol status, disparity consistency, and word alignment using periodic K28.5 control characters. This preserves bounded disparity and transition activity for link analysis without claiming to reproduce the FEB firmware or CDR loop. Each eight-bit item is represented by ten serial bits, so Equation 2.12 relates the raw 8b10b payload capacity $R_{payload}$ to the serial line rate $R_{line}$ which is 1.25 Gbps for the MuPix11 link considered here. This coded-link model evaluates 8b10b symbol integrity and projected checker events; its stated payload rate follows the explicitly defined diagnostic framing rather than an exact reproduction of the MuPix firmware packet scheduler or control/idle pattern [11].

$$R_{payload} = \frac{8}{10} R_{line} \leftrightarrow R_{line} = \frac{10}{8} R_{payload} \qquad (2.12)$$

For the diagnostic record, MATLAB encodes 512 payload bytes and inserts one K28.5 control group before every 64 payload bytes. The resulting eight K28.5 groups and 512 data groups form 520 ten-bit groups; at a 1.25 Gbps line rate, this diagnostic framing has a 0.9846 Gbps payload rate.

For the independent-error projection, MATLAB enumerates every received ten-bit word by Hamming weight against all legal data and K28.5 words in both running-disparity states. In Equation 2.13, $R_{line}/10$ is the transmitted ten-bit group rate, $N_{tx}$ is the number of enumerated legal transmitted words, $p$ is the raw independent bit-error probability, and $N_{chk,w}$ counts checker-triggering received outcomes at Hamming weight $w$. The bracketed sum is therefore the average immediate checker-event probability per transmitted ten-bit group, and multiplication by the group rate gives $R_{chk}$ in events/s. Valid-symbol-mismatch and logical-union combined-event rates are calculated separately. These are model projections: burst correlation, CDR loss, comma/framing loss, and later running-disparity propagation are outside the model.

$$R_{chk} = \frac{R_{line}}{10}\left[\frac{1}{N_{tx}}\sum_{w=0}^{10} N_{chk,w}\, p^{w}(1-p)^{10-w}\right] \tag{2.13}$$

Separately, the finite receiver record contains 32,749 aligned serial decisions. If zero errors are observed, the 99% upper bound follows Equation 2.11; it is a finite-record confidence bound, not a measured physical BER or a replacement for the projected checker-event rate.

## 3. Probe layout and PTSL CST Studio extracts

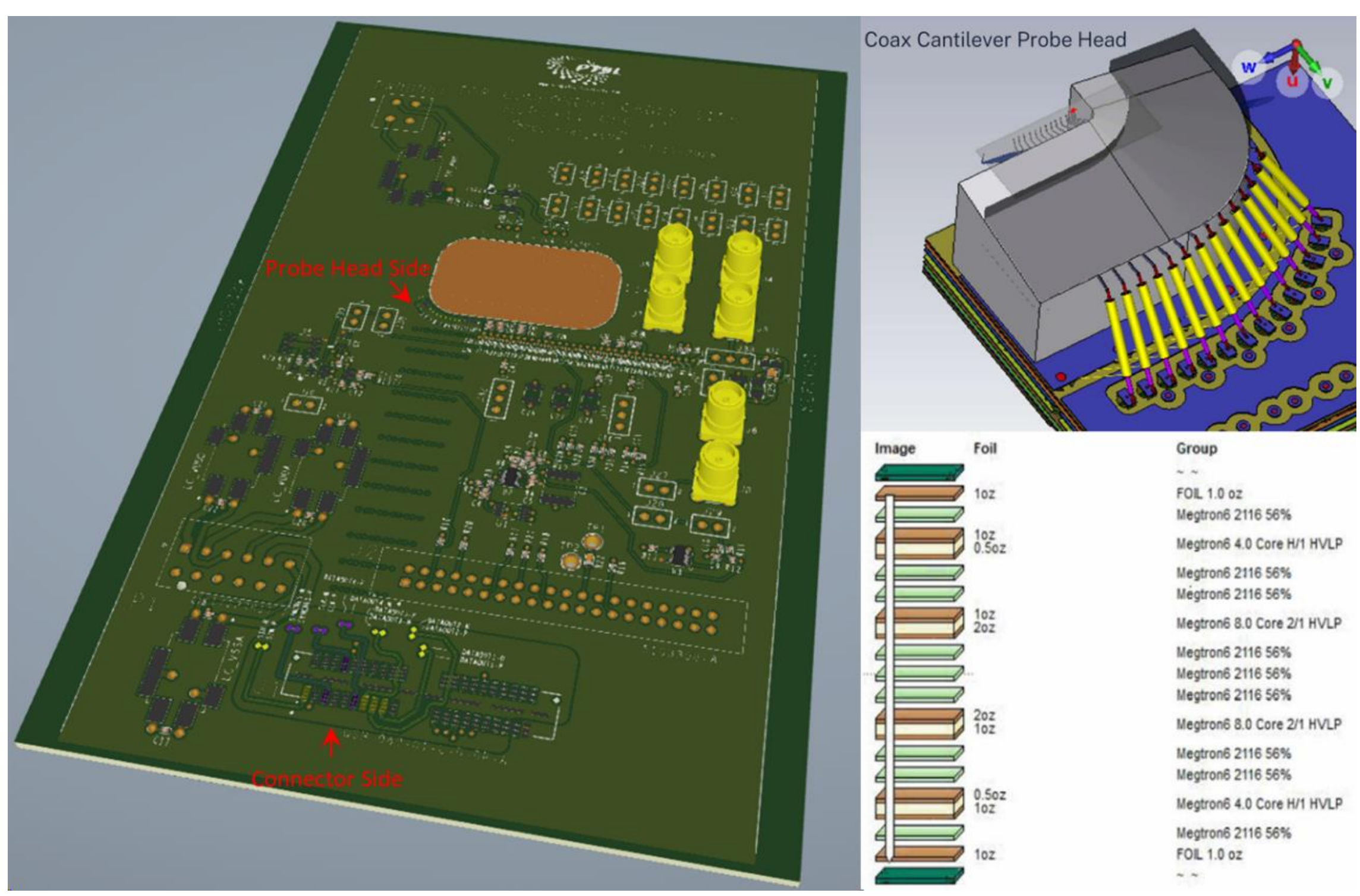

Figure 2. PTSL probe-card overview.

We oversaw the PTSL probe-card layout from the initial design concept to the design for manufacturing (DFM) stage. The design review identified localized IPC-2223 compliance concerns regarding connector anti-pad sizes and via annular-ring geometries. While these enlarged anti-pad clearances suppress parasitic capacitance to maintain target differential impedance, they simultaneously introduce return-current discontinuities across the reference plane. The Oxford MATLAB analysis framework evaluates the impact of these discontinuities on pre-production performance using the supplied S-parameters, though final verification still requires empirical hardware validation. Figure 2 shows the 3D layout.

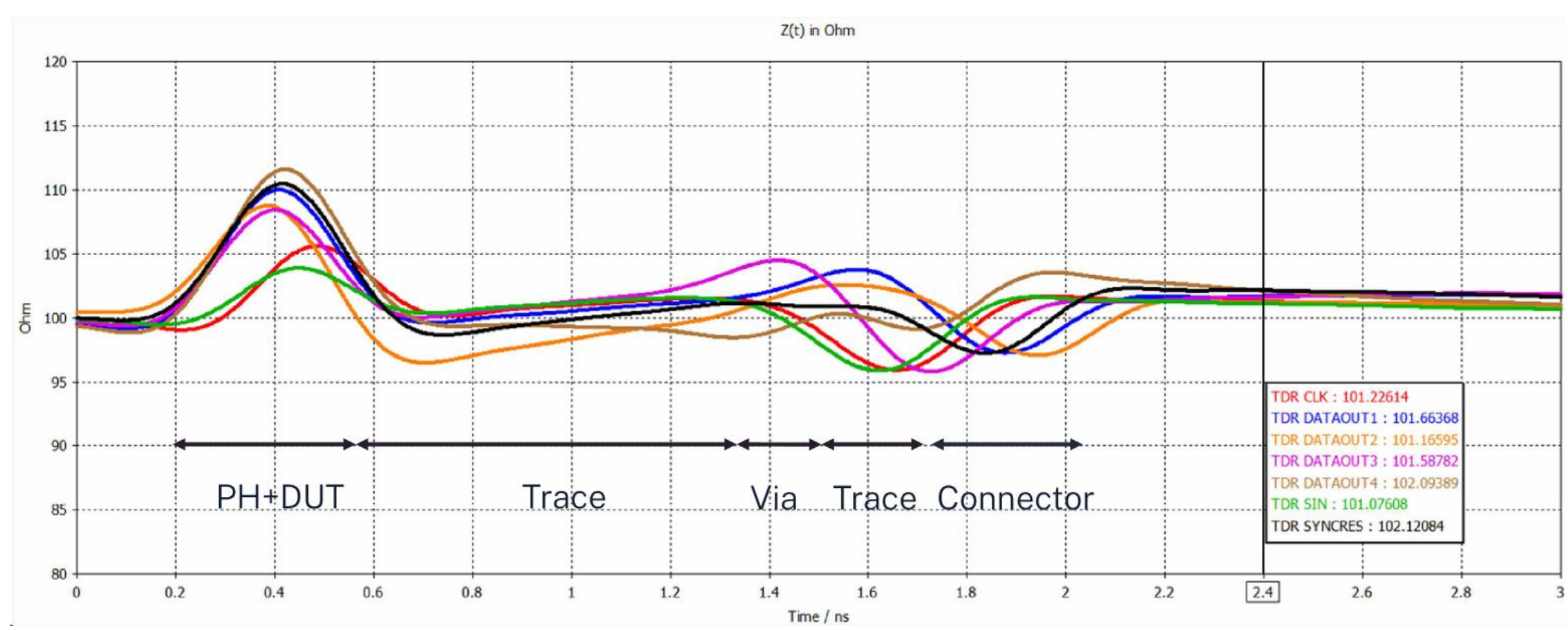

Figure 3. CST time-domain reflectometry profiles for all probe-card differential pairs.

Table 1. PTSL-reported route geometry and CST differential S-parameter magnitudes at 1.250 GHz [5]. DATAOUT1–DATAOUT4 correspond to high-speed DP1–DP4 links.

| No. | Diff. pair | Signal | Length per segment | | | Total physical length (mm) | PH RL | PH → Connector IL | Connector RL |
|---|---|---|---|---|---|---|---|---|---|
| | | | Cantilever RF probe (mm) | PCB trace length (mm) | Connector pin (mm) | | SDD11 (dB) | SDD21 (dB) | SDD22 (dB) |
| 1 | CLK | CLK_N | 10.85 | 81.70 | 3.43 | 95.97 | −18.93 | −0.55 | −18.04 |
| | | CLK_P | 10.85 | 81.70 | 3.43 | 95.98 | | | |
| 2 | SIN | SIN_N | 10.85 | 98.07 | 3.43 | 112.34 | −20.35 | −0.52 | −19.06 |
| | | SIN_P | 10.85 | 98.08 | 3.43 | 112.35 | | | |
| 3 | DATA OUT1 | DATAOUT1_N | 10.85 | 99.90 | 3.43 | 114.17 | −22.27 | −0.56 | −22.34 |
| | | DATAOUT1_P | 10.85 | 99.91 | 3.43 | 114.18 | | | |
| 4 | DATA OUT2 | DATAOUT2_N | 10.85 | 109.05 | 3.43 | 123.33 | −26.31 | −0.54 | −28.59 |
| | | DATAOUT2_P | 10.85 | 109.05 | 3.43 | 123.32 | | | |
| 5 | DATA OUT3 | DATAOUT3_N | 10.85 | 86.57 | 3.43 | 100.84 | −17.44 | −0.54 | −16.72 |
| | | DATAOUT3_P | 10.85 | 86.57 | 3.43 | 100.84 | | | |
| 6 | DATA OUT4 | DATAOUT4_N | 10.85 | 85.13 | 3.43 | 99.41 | −17.34 | −0.54 | −17.22 |
| | | DATAOUT4_P | 10.85 | 85.16 | 3.43 | 99.43 | | | |
| 7 | SYNC RES | SYNCRES_N | 10.85 | 98.50 | 3.43 | 112.77 | −18.57 | −0.57 | −18.44 |
| | | SYNCRES_P | 10.85 | 98.49 | 3.43 | 112.77 | | | |

The design targets signal-integrity performance up to 1.250 GHz—twice the 0.625 GHz Nyquist frequency of the 1.25 Gbps NRZ data stream. This twofold-Nyquist frequency is a spectral design-robustness check: it tests S-parameter margin beyond the receiver operating point, but does not claim a validated 2.5 Gbps link rate. At 1.250 GHz, insertion loss (SDD21) is specified to be no worse than −1.5 dB, with probe-head return loss (SDD11) kept at or below −15 dB; the SDD21 and SDD11 magnitude curves should not cross below 3.750 GHz, the third harmonic of the 1.250 GHz design-check frequency. Baseline CST Studio-derived S-parameters and virtual TDR simulations supplied by PTSL [5] are presented first, ensuring raw electromagnetic evidence remains clearly distinguished from derived link-level pipeline inferences.

Figure 3 shows the PTSL differential time-domain reflectometry (TDR) profiles. The traces remain close to the 100 Ω target through the PCB sections, with localized impedance excursions at the probe head (PH), vias, and connector transitions. The connector and routing-via anti-pad geometries influence these departures. Table 1 reports the PTSL R03 full-wave CST quantities at 1.250 GHz, with the probe head defined as the excitation side and the connector as the response side. SDD11 is the probe-head return loss (PH RL), SDD21 is the differential insertion loss from PH to connector (PH → Connector IL), and SDD22 is the connector return loss (Connector RL). In the R03 summary table, SDD21 ranges from −0.57 to −0.52 dB, SDD11 ranges from −26.31 to −17.34 dB, and SDD22 ranges from −28.59 to −16.72 dB; all channels satisfy the stated −1.5 dB insertion-loss and −15 dB return-loss criteria at 1.250 GHz. The fixed-port MATLAB recomputation using the supplied Touchstone files is presented separately in Table 5 at the same 1.250 GHz comparison frequency.

## 4. Signal-path results

The PTSL-exported ODB++ layout was imported into the CST Studio 3D electromagnetic field solver with port configurations of excitation (1,2) → response (3,4). Single-ended S-parameters were reanalyzed using the Oxford MATLAB-based framework as discussed in Section 2. Among four high-speed channels, the DP4 channel's findings are narrated here.

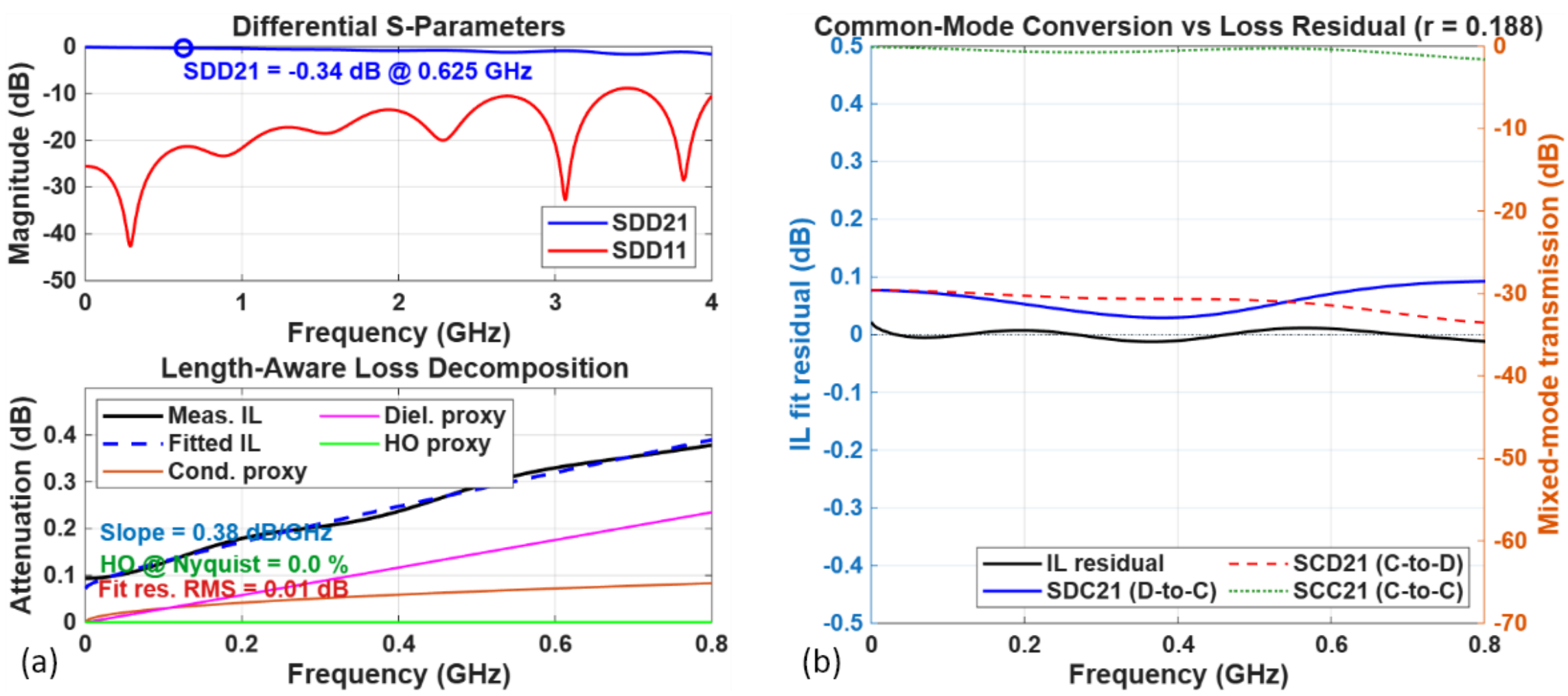

Figure 4. (a) Differential S-parameters and the length-aware operating-band loss decomposition (10 MHz to 800 MHz) with different proxies; (b) mixed-mode conversion and loss-residual correlation (10 MHz to 800 MHz).

Figure 4(a) establishes differential insertion/return loss, operating Nyquist behavior, measured-band headroom, and a physically motivated decomposition of smooth attenuation. The non-negative fit uses $\sqrt{f}$ conductor, $f$ dielectric, and $f^2/f^4$ residual-curvature bases (higher-order (HO) proxy), normalized by the exact 99.42 mm DP4 path. The fixture constant remains a whole-path intercept; it is not divided into imaginary material segments. As shown, the IL (SDD21) of DP4 is −0.34 dB at 1.25 Gbps (Nyquist: 0.625 GHz). For the 0.01–0.80 GHz fit, the conductor-loss proxy and dielectric-loss proxy are 0.073 dB and 0.183 dB, respectively, at 0.625 GHz. The non-crossing condition between SDD21 and SDD11 up to 3.750 GHz—the third harmonic of the 1.250 GHz design-check frequency—indicates expected negligible bounce-back noise dominance. The higher-order residual proxy is 0 dB for this constrained fit. In other words, after the constant, $\sqrt{f}$, and $f$ bases are fitted, the 0.01–0.80 GHz DP4 data do not require an additional $f^2/f^4$ curvature term. This does not demonstrate that the physical route radiates no power. Together with the whole-path intercept, the 0.073 dB conductor-loss contribution and 0.183 dB dielectric-loss contribution quantitatively reconstruct the smooth DP4 loss within the declared model. They are not separately measured copper or dielectric losses. The correlated bases permit more than one allocation of the same smooth attenuation, so additional coupon, material, or de-embedded fixture information would be needed only to attribute loss to a specific physical mechanism.

A separate DP4 audit refit over 0.01–1.50 GHz (not shown) yields a 0.010 dB RMS residual and reconstructs the 0.625 GHz CST insertion-loss magnitude to 0.324 dB, compared with 0.336 dB. The conductor/dielectric/higher-order allocation moves from 0.073/0.183/0.000 dB to 0.099/0.154/0.004 dB. These remain quantitative contributions of their respective declared fits, while their allocation is fitting-band dependent and is not a set of independent material or radiation measurements.

Figure 4(b) reports the quantifiable deterministic conversion between differential and common modes and tests whether the conversion structure tracks unexplained differential-loss residuals. The assessed single-ended matrix is transformed with power-normalized sum/difference waves. Following the plot's excitation→response naming: displayed SDC21 = −29.69 dB is D→C, displayed SCD21 = −31.695 dB is C→D, SCC21 = −0.52 dB is C→C, and D→C relative = −29.354 dBc. SCC21 = −0.52 dB indicates low end-to-end common-mode attenuation through the assessed probe-head-to-connector path, consistent with the absence of a major common-mode discontinuity in the modeled interconnect. This aggregate result does not isolate individual needle, via, or connector contributions, nor does it quantify common-to-differential conversion, which is assessed separately by SCD21. Here, SCD21 = −31.695 dB indicates weak common-to-differential conversion at the response side. The D→C and C→D conversion terms are distinct physical directions and must not be interchanged. Although the insertion-loss residual and modal-conversion traces show a weak correlation (r = 0.1876), this reflects shared frequency dependence rather than deterministic power loss via mode conversion.

A MuPix-like differential signal is launched with a deterministic PRBS-31 generator that maps the generated bits to NRZ levels of plus or minus one. The one-UI post-cursor and causal first-order edge are defined below. The waveform is normalized to 0.7 $V_{pp}$ differential and propagated through the rational approximation of the loaded channel transfer function. The TX eye (Figure 5 (a)) establishes the launch reference prior to the simulated passive channel. Eye height quantifies vertical separation at the selected sampling phase; DDJ crossing spread measures pattern-dependent crossing-time variation. The eye-analysis routine reshapes the finite-edge, post-cursor transmitter waveform into two-UI rows, uses the 5th and 95th percentiles for phase search, and uses the 1st and 99th percentiles for the reported opening. The crossing-analysis routine linearly interpolates zero crossings, tests the equivalent half-UI crossing clusters, selects the best-supported cluster, and reports its 1st-to-99th-percentile width. A display-only circular shift centers the eye; it is not used for BER. TX eye height is 0.596 V and DDJ crossing spread is 0.008 UI, or about 6.4 ps at the 800 ps UI. The nominal electrical launch setting is 0.7 $V_{pp}$ differential, with an 180 ps rise time and a −0.08 post-cursor.

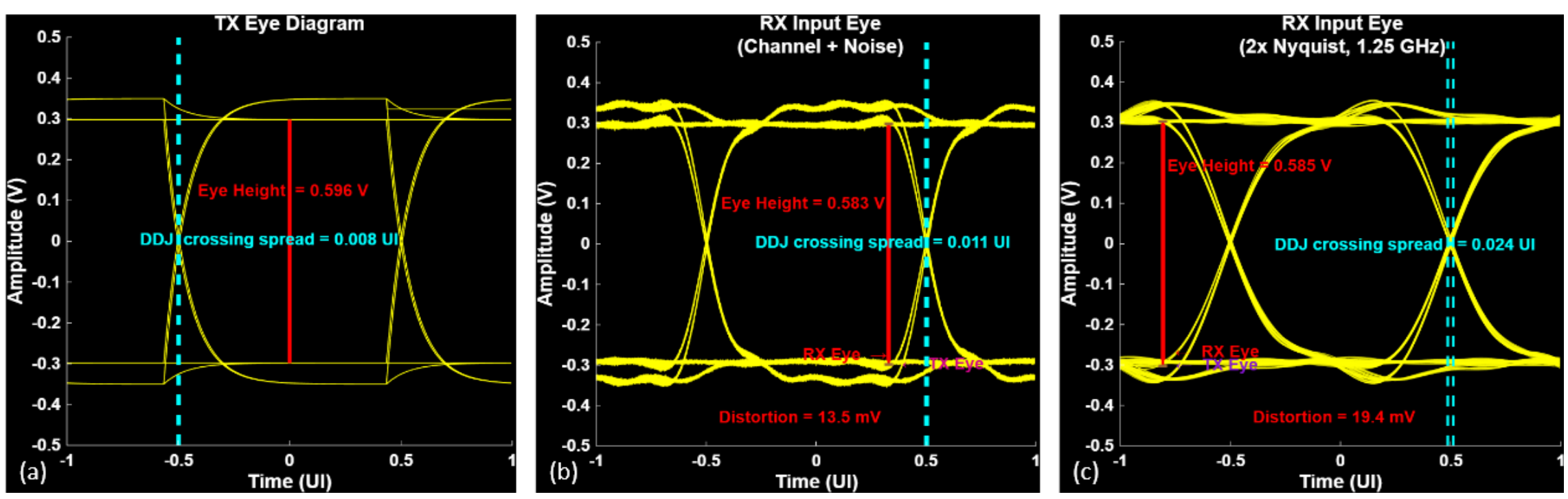

Figure 5. (a) Modeled MuPix-side transmitter eye; (b) FEB receiver-input eye with nominal input noise of 2 mV (RMS), aperture jitter of 2 ps (RMS) for BER, and (c) modeled FEB receiver-input eye at 2× Nyquist (1.25 GHz). The eye diagram at 1.25 GHz is a spectral-stress condition, not a demonstrated or validated 2.5 Gbps operating point.

The received waveform on the Front-End Board (FEB) is obtained by convolving the transmitter waveform with the loaded channel impulse response. Source and load mismatch are retained through the loaded two-port expression below; therefore, both magnitude and phase influence the pulse response. In the Oxford analysis framework, it first finds the maximum-opening receiver phase, aligns only the plotted eye, and recenters the dominant crossing cluster. The eye-analysis routine returns the 1st/99th-percentile voltage limits, and the crossing-analysis routine returns the corresponding transition bounds. No additive receiver noise, Continuous-Time Linear Equalizer (CTLE), or Decision Feedback Equalizer (DFE) is present, so the figure isolates the measured channel plus the stated source/load boundary model. At 1.25 Gbps, the eye height remains 0.593 V. Even with the inclusion of the 2 mV RMS random input noise and 2 ps flat-top aperture jitter, the operating point remains limited to ~13 mV (Figure 5(b)). Panel 5(c) uses the 2×Nyquist, 1.25 GHz spectral-stress condition; its ~19.4 mV distortion is a model stress result, not a demonstrated or validated 2.5 Gbps link operating point.

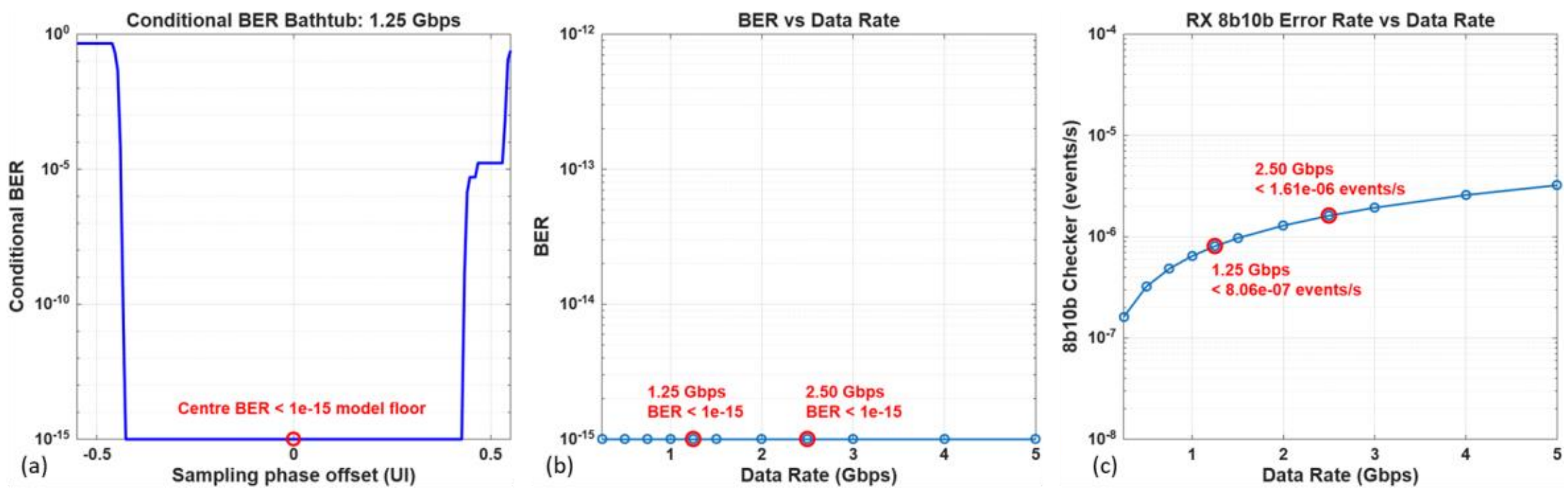

Figure 6. (a) Receiver BER bathtub; (b) receiver BER versus data rate; (c) projected receiver 8b10b checker/event rate versus data rate.

Figure 6(a) shows the BER bathtub, which shifts the receiver sampling phase across the UI while preserving channel ISI and the configured noise/jitter model. Its minimum identifies the best phase; the slopes indicate sensitivity to clock placement and deterministic timing closure. For each phase, the deterministic context voltage and local slope define a Gaussian decision tail. The DP4 timing scan uses 161 phase positions from −0.55 to +0.55 UI, with a 0.006875-UI increment, and linear interpolation of the 200-sample/UI waveform. At each phase, the raw retained probability is the maximum of the context-resolved Gaussian tail, the robust median level-spread guard, and the deterministic slicer-error fraction. The displayed wall then uses the one-sided cumulative maximum, as in Eq. (4.1), moving outwards from the recovered center c.

$$B_{env}(\varphi_i) = \begin{cases} \max\limits_{i \le j \le c} B_{raw}(\varphi_j) & for\ i \le c, \\ \max\limits_{c \le j \le i} B_{raw}(\varphi_j) & for\ i \ge c. \end{cases} \quad (4.1)$$

Here, $\varphi_i$ represents the discrete sampling phase offset at step i, $\varphi_c$ (indexed at c) is the recovered optimal phase center of the eye, $B_{raw}(\varphi_j)$ is the context-resolved BER evaluated at phase step j, and $B_{env}(\varphi_i)$ constructs a monotonic BER envelope moving outward from the center. This cumulative operation ensures that clock-phase offsets are evaluated against worst-case edge margins, preventing raw numerical fluctuations or local statistical crossovers from producing non-physical decreases in BER as the sampling point shifts away from the eye center.

At the recovered DP4 center, the context tail is 2.225e−308, and the robust level-spread term is 2.281e−297; the plot shows these levels $10^{-15}$ as unresolved beneath a declared credibility floor. The conditional-one/zero ISI spreads are 7.76/7.73 mV, and their slope-weighted jitter terms are 0.17/0.17 mV. A shallow plateau on a wall therefore indicates that the running maximum has held a previous raw estimate, commonly near a component crossover or one-error quantization step of approximately 3.054e−05. It is not a downward BER excursion and cannot, without raw-component and perturbation checks, be assigned to a reflection or resonance.

Every DP4 rate uses the same full-aligned PRBS decision definition; no low- and high-rate models are spliced. Sub-floor points in Figure 6(b) are displayed at rather than at numerical underflow, so the 1.25 Gbps marker means only that this model does not resolve a larger BER. It is not a zero-error hardware measurement.

In Figure 6(c), the DP4 receiver checker enumerates immediate ten-bit groups over all 1024 possible received words and retains invalid-code, disparity, and valid-symbol-mismatch classes separately. Its event rate is a projection based on model probabilities, not a measured count. Burst memory, clock-data recovery, framing loss, and later running-disparity propagation remain outside the enumeration. At the operating point, the checker-event projection is below the 8.064e−07 events/s reporting bound, while the zero-count 99% upper bound is 1.111e+06 events/s.

## 5. Comparative analysis

Table 2 is the route-resolved evidence: it preserves each operating-point quantity before any averaging or subtraction. All four routes include the same 10.846 mm probe and 3.429 mm connector-pin sections, so the 99.420–123.325 mm path range comes entirely from the PCB trace. Despite the 23.905 mm span, SDD21 occupies only −0.35 to −0.30 dB, and the channel-only eye occupies 0.586 to 0.588 V. The tighter transmission and eye ranges must not conceal the larger modal differences: D-to-C spans −31.038 to −28.236 dBc, while SCC21 spans −1.244 to −0.52 dB. The $\sqrt{f}$ conductor-loss proxy, $f$ dielectric-loss proxy, and higher-order residual proxy columns report the non-negative contributions assigned by the declared 0.01–0.80 GHz attenuation fit. Together with the fitted constant intercept, their sum reconstructs the smooth SDD21 attenuation over that band. Because these fitted basis functions are correlated, the allocation among these terms is model-dependent and should not be interpreted as independently measured copper or dielectric losses for an individual route.

The columns answer distinct questions. SDD21 is the wanted differential transfer; D-to-C is the conversion of a differential excitation into common mode; SCC21 is the forward transfer of a common-mode excitation already present at the input; and eye height is the time-domain consequence of usable-band magnitude, phase, and pattern response under the common transmitter and receiver assumptions. Neither D-to-C nor SCC21 is random noise, and one cannot be substituted for the other.

Table 2. DP1-DP4 route-specific operating-point quantities at 1.25 Gbps (0.625 GHz Nyquist). The reported loss proxies belong to the declared 0.01–0.80 GHz non-negative attenuation fit.

| DP | Path (mm) | SDD21 (dB) | Cond. proxy (dB) | Diel. proxy (dB) | Higher proxy (dB) | D-to-C dBc | SCC21 dB | RX eye V |
|---|---|---|---|---|---|---|---|---|
| DP1 | 114.175 | −0.340 | 0.303 | 0.000 | 0.020 | −31.038 | −1.028 | 0.586 |
| DP2 | 123.325 | −0.350 | 0.293 | 0.000 | 0.021 | −28.236 | −1.244 | 0.586 |
| DP3 | 100.840 | −0.300 | 0.261 | 0.000 | 0.003 | −29.242 | −0.531 | 0.587 |
| DP4 | 99.420 | −0.340 | 0.073 | 0.183 | 0.000 | −29.354 | −0.520 | 0.588 |

Table 3. DP1-DP4 descriptive statistics; dispersion is the sample standard deviation (SD).

| Metric | Mean | Sample SD | Minimum | Maximum |
|---|---|---|---|---|
| Path length (mm) | 109.4400 | 11.3955 | 99.4200 | 123.3250 |
| SDD21 (dB) | −0.3325 | 0.0222 | −0.3500 | −0.3000 |
| Loss slope (dB/GHz) | 0.3625 | 0.0419 | 0.3000 | 0.3900 |
| Conductor contribution (dB) | 0.2325 | 0.1078 | 0.0730 | 0.3030 |
| Dielectric proxy (dB) | 0.0457 | 0.0915 | 0.0000 | 0.1830 |
| Higher-order proxy (dB) | 0.0110 | 0.0110 | 0.0000 | 0.0210 |
| D-to-C relative (dBc) | −29.4675 | 1.1614 | −31.0380 | −28.2360 |
| SCC21 (dB) | −0.8307 | 0.3634 | −1.2440 | −0.5200 |
| RX eye (V) | 0.5867 | 0.0010 | 0.5860 | 0.5880 |
| Fit residual RMS (dB) | 0.0063 | 0.0010 | 0.0050 | 0.0070 |

Table 4. Signed route differences relative to DP4 and absolute BER/8b10b combined-event-rate upper bounds at 1.25 Gbps (Nyquist 0.625 GHz). Because every route is below the reporting floor, the final two columns are common bounds and are not route-ranking metrics.

| Route | Path diff. vs DP4 (mm) | SDD21 diff. vs DP4 (dB) | D-to-C diff. vs DP4 (dB) | SCC21 diff. vs DP4 (dB) | RX-eye diff. vs DP4 (mV) | FEB-eye diff. vs DP4 (mV) | Analytical BER (upper bound) | Projected 8b10b combined event rate (upper bound, events/s) |
|---|---|---|---|---|---|---|---|---|
| DP1 | +14.755 | +0.000 | −1.684 | −0.508 | −2.0 | −3.0 | $< 10^{-15}$ | $< 1.25 \times 10^{-6}$ |
| DP2 | +23.905 | −0.010 | +1.118 | −0.724 | −2.0 | −3.0 | $< 10^{-15}$ | $< 1.25 \times 10^{-6}$ |
| DP3 | +1.420 | +0.040 | +0.112 | −0.011 | −1.0 | −1.0 | $< 10^{-15}$ | $< 1.25 \times 10^{-6}$ |
| DP4 (ref.) | 0.000 | 0.000 | 0.000 | 0.000 | 0.0 | 0.0 | $< 10^{-15}$ | $< 1.25 \times 10^{-6}$ |

Table 3 compresses the four absolute rows into sample-level location, spread, and range. It shows whether a metric is tightly clustered or route-sensitive; its mean is not a replacement for a trace result, and its sample standard deviation is neither a fabrication tolerance nor a measurement uncertainty. Table 4 therefore returns from population description to paired engineering contrasts. DP4 is selected as the reference because it is the shortest route and, under the common time-domain model, has the largest channel-only and FEB-input eyes and the smallest reported channel-induced vertical closure. It is also worth noting that, among the MuPix11 DP1–DP4 data-pair channels on the ladder, only DP4 is spTAB-bonded to the outer-pixel-detector HDI-flex [12]. This baseline choice does not imply that DP4 is superior in insertion loss or modal isolation.

DP1 is 14.755 mm longer than DP4, yet both report SDD21 = −0.34 dB at 0.625 GHz. This equality at one frequency does not establish equal broadband loss. DP1's D-to-C value is 1.684 dB more negative; expressed as a power ratio, its converted fraction is 0.679 times that of DP4, or about 32.1% lower. Conversely, DP1's SCC21 is 0.508 dB lower, so it transmits less of an already-common-mode incident wave. Its channel-only and FEB-input eyes are only 2.0 and 3.0 mV below DP4. The comparison separates a meaningful modal-isolation advantage from a very small vertical-eye difference.

DP2 adds 23.905 mm of path relative to DP4 but changes Nyquist SDD21 by only −0.010 dB; its channel-only and FEB-input eyes are lower by only 2.0 and 3.0 mV. The more discriminating result is D-to-C = −28.236 dBc, 1.118 dB less isolated than DP4 and equivalent to approximately 1.294 times its converted-power fraction. DP2 also has the largest pattern-conditioned zero/one-level spreads (11.65/11.97 mV). Those spreads are deterministic context dispersion under the stated model, not additive noise amplitudes. The result shows that extra route length alone does not predict modal balance, even when the final vertical opening remains almost unchanged.

DP3 is only 1.420 mm longer than DP4 and therefore provides the closest geometric control. It has SDD21 = −0.3 dB, +0.040 dB less negative than DP4, but its channel-only and FEB-input eyes are still 1.0 mV smaller. Its D-to-C difference is only +0.112 dB (1.026 times the DP4 converted-power fraction), and SCC21 differs by −0.011 dB. Thus, DP3 and DP4 are nearly modal-equivalent at the operating point, while their single-frequency differential transfer and integrated eye ranking are not identical.

The non-negative loss fit assigns DP3 0.261/0.000/0.003 dB and DP4 0.073/0.183/0.000 dB to the conductor-loss/dielectric-loss/higher-order residual contributions of the declared model. After path normalization, these become 2.588/0.000/0.025 dB/m for DP3 and 0.739/1.841/0.000 dB/m for DP4. Although the residual RMS values are only 0.005 and 0.007 dB, redistribution among correlated non-negative basis functions changes the allocation without invalidating the reconstructed total loss; it does not establish that adjacent traces contain different copper or dielectric materials.

In voltage-ratio terms, DP1/DP2/DP3/DP4 transmit approximately 0.888/0.867/0.941/0.942 of an already-common-mode incident wave. A value nearer 0 dB therefore means stronger common-to-common propagation, not stronger differential-to-common conversion. DP3 and DP4 pass an imposed common-mode component more efficiently than DP1 and DP2, while DP1 provides the strongest D-to-C isolation. These independent modal properties require separate acceptance criteria.

The channel-only eye varies by only 2.0 mV across all four routes, and channel-induced closure improves from 10.51 mV on DP2 to 8.32 mV on DP4. This supports a robust modeled vertical margin for every short route, but does not make their mode-conversion behavior equivalent or validate sub-$10^{-15}$ analytical tails as measured BER. The ranking is therefore metric-dependent: DP2 is longest, DP4 is shortest and has the largest reported eye, DP3 has the least-negative Nyquist SDD21, and DP1 has the strongest D-to-C isolation at −31.038 dBc. No single trace is 'best' unless the acceptance quantity is stated explicitly.

The final two columns of Table 4 are absolute bounds, not differences. DP1-DP4 each have an analytical BER below $10^{-15}$ at 1.25 Gbps and a corresponding 8b10b combined-event-rate upper bound below $1.25 \times 10^{-6}$ events/s. Their equality follows from the common receiver assumptions and the fact that all four analytical BER results lie beneath the declared reporting floor. Reporting these values as inequalities avoids attributing physical meaning to the raw floating-point tail; the latter is retained only in the unaggregated per-route MATLAB text logs for software traceability.

The 8b10b combined-event bound is likewise a model projection rather than an observed event rate. The combined class is the logical union of a checker event (invalid code or running-disparity violation) and a valid-symbol mismatch; it is therefore the appropriate end-to-end coded-link metric. The associated checker-only bound is below 8.064e−07 events/s. Immediate ten-bit-word enumeration uses the same 1.25 Gbps line rate and exhaustive 1024-word weighting for every route, with burst memory, clock-data recovery, framing loss, and subsequent running-disparity propagation outside the model. In the finite PRBS record, zero errors among 32749 aligned decisions support only a 99% BER upper bound of 1.406e−04; the encoded zero-count record gives a 99% checker-rate upper bound of 1.111e+06 events/s. Consequently, Table 4 supports no BER or coded-event ranking among DP1-DP4. The defensible manuscript statement remains that the analytical BER is below $10^{-15}$, pending a longer independent simulation or timed hardware measurement.

Across DP1-DP4, the mean physical route is 109.44 mm, and the modeled FEB-input eye is 0.5867 ± 0.001 V (mean ± sample standard deviation) at 1.25 Gbps. The 23.905 mm path-length span produces only a 0.05 dB range in SDD21 and a 2 mV range in the channel-only eye, so all four routes retain favorable modeled differential margin. The comparative analysis nevertheless shows that route length is not a sufficient ranking variable: DP1 has the strongest differential-to-common isolation (−31.038 dBc), DP2 is the longest and has the weakest isolation (−28.236 dBc), DP3 has the least-negative Nyquist SDD21 (−0.3 dB), and the shortest route DP4 has the largest reported eye. The design choice must therefore name the acceptance metric rather than assign a single overall 'best' route.

Table 5. MATLAB fixed-port, power-normalized mixed-mode recomputation at 1.250 GHz for direct comparison with Table 1. This frequency is twice the 0.625 GHz Nyquist frequency of the 1.25 Gbps NRZ link and is used only as a spectral robustness cross-check, not as a 2.5 Gbps receiver or BER operating point.

| Differential pair | PH RL SDD11 (dB) | PH → Connector IL SDD21 (dB) | Connector RL SDD22 (dB) |
|---|---|---|---|
| CLK | −18.93 | −0.55 | −18.05 |
| SIN | −20.36 | −0.53 | −19.07 |
| DP1 | −22.27 | −0.57 | −22.34 |
| DP2 | −26.32 | −0.54 | −28.59 |
| DP3 | −17.44 | −0.55 | −16.73 |
| DP4 | −17.34 | −0.55 | −17.23 |
| SYNCRES | −18.58 | −0.58 | −18.44 |

At the common 1.250 GHz comparison frequency, the MATLAB values in Table 5 reproduce the published two-decimal PTSL R03 summary values in Table 1, with a largest displayed difference of 0.01 dB. Comparison with the unrounded R03 quantities is within 0.005 dB. Together, these results confirm the fixed-port mixed-mode transformation as a twofold Nyquist spectral robustness cross-check of the PTSL source model; they do not demonstrate 2.5 Gbps receiver eye or BER performance. The calculation uses the Touchstone header port pairs (1,2) at the probe head and (3,4) at the connector, followed by the power-normalized mixed-mode transform. The small displayed differences arise from tabulated rounding, interpolation of sampled complex S-parameters to 1.250 GHz, and numerical precision; they do not indicate a physical discrepancy. This is a source-model cross-check, not an independent validation of the complete receiver and BER chain, because the MATLAB analysis uses the PTSL-exported Touchstone data. Independent validation of the complete MATLAB chain requires either a calibrated and de-embedded VNA/hardware measurement or a separate benchmark channel, and this remains future work.

## 6. Conclusions

The principal contribution of this work is a reusable MATLAB signal-integrity framework that turns compatible four-port Touchstone data into traceable link-level evidence. It combines checked port pairing, power-normalized mixed-mode conversion, a causal loaded-channel representation, complete-record waveform propagation, receiver decision analysis, and 8b10b diagnostic projections into a single reproducible chain. Crucially, the workflow keeps physical channel behavior, declared transmitter and receiver assumptions, analytical model projections, and finite-record observations separate. Accordingly, its value lies in providing a reproducible evidence trail from checked four-port inputs to separately auditable channel, receiver, and protocol-level results.

The PPCB-1347-MuPix11 probe card offers a focused case study of what this framework reveals beyond a single loss metric. At 1.25 Gbps, the four modeled links retain FEB-input eyes of 0.5867 ± 0.001 V and exhibit only a 0.05 dB Nyquist-SDD21 span under the stated assumptions. Yet their comparison is metric-dependent: DP3 has the least-negative Nyquist insertion loss, DP1 the strongest differential-to-common isolation, and DP4 the largest modeled eye. Physical length alone is therefore not an adequate proxy for link quality; acceptance criteria must specify whether they control loss, modal conversion, eye margin, or another receiver-relevant quantity. The same framework can analyze compatible four-port responses from electromagnetic simulation or calibrated, de-embedded VNA measurements while retaining visible, configurable data rate, transmitter, receiver, and coding assumptions. The present Table 5 comparison cross-checks the MATLAB mixed-mode transformation against the PTSL R03 summary data and its CST/Touchstone source model; it does not independently validate the complete waveform and BER chain. Calibrated VNA/hardware measurements or a separate benchmark channel are therefore the appropriate next step in validation. Until then, sub-floor BER and coded-event results are reported as analytical bounds rather than hardware measurements.

## Data Availability Statement

The four-port Touchstone data, virtual TDR results, and route geometry analyzed in this work were supplied by PTSL and are documented in the internal technical report cited in [5]. They are not publicly available because they contain proprietary pre-production probe-card design information. Requests for access to these materials should be directed to the corresponding author and are subject to PTSL approval.

## Ethics Statement

This work involved no human participants, animals, personal data, or clinical material. Ethics approval and informed consent were therefore not required.

## Acknowledgments

The project has been funded by the Science and Technology Research Council, UK (STAK00085). The authors thank the PTSL RF team for sharing the PPCB-1347-MuPix11 signal-integrity and layout inputs for this analysis, and the Mu3e collaboration for fruitful discussions of the signal-integrity findings.

## Conflicts of Interest

The authors declare that there are no conflicts of interest.